\documentclass[twocolumn]{aastex63}

\newcommand{\chn}{{\it Chandra}}

\usepackage{graphicx}
\usepackage{longtable}
\usepackage{mdwlist}

\usepackage{natbib}
\usepackage{multirow}
\usepackage{upgreek}
\usepackage{wasysym}
\usepackage{txfonts}
\usepackage{gensymb}
\usepackage{morefloats}
\usepackage{threeparttable}
\usepackage{rotating}
\usepackage{graphicx}
\usepackage{color}

\shorttitle{Optical and X-ray filaments in 3CR\,318.1}
\shortauthors{Jimenez Gallardo et al.}

\begin{document}
\title{Raining in MKW\,3s: a Chandra-MUSE analysis of X-ray cold filaments around 3CR\,318.1} 
\correspondingauthor{Ana Jimenez-Gallardo}
\email{ana.jimenezgallardo@edu.unito.it}

\author[0000-0003-4413-7722]{A. Jimenez-Gallardo}
\affiliation{Dipartimento di Fisica, Universit\`a degli Studi di Torino, via Pietro Giuria 1, I-10125 Torino, Italy}
\affiliation{European Southern Observatory, Alonso de C\'ordova 3107, Vitacura, Regi\'on Metropolitana, Chile}
\affiliation{Istituto Nazionale di Fisica Nucleare, Sezione di Torino, I-10125 Torino, Italy}
\affiliation{INAF-Osservatorio Astrofisico di Torino, via Osservatorio 20, 10025 Pino Torinese, Italy}

\author[0000-0002-1704-9850]{F. Massaro}
\affiliation{Dipartimento di Fisica, Universit\`a degli Studi di Torino, via Pietro Giuria 1, I-10125 Torino, Italy}
\affiliation{Istituto Nazionale di Fisica Nucleare, Sezione di Torino, I-10125 Torino, Italy}
\affiliation{INAF-Osservatorio Astrofisico di Torino, via Osservatorio 20, 10025 Pino Torinese, Italy}
\affiliation{Consorzio Interuniversitario per la Fisica Spaziale, via Pietro Giuria 1, I-10125 Torino, Italy}

\author[0000-0002-0690-0638]{B. Balmaverde}
\affiliation{INAF-Osservatorio Astrofisico di Torino, via Osservatorio 20, 10025 Pino Torinese, Italy}

\author[0000-0002-5646-2410]{A. Paggi}
\affiliation{Dipartimento di Fisica, Universit\`a degli Studi di Torino, via Pietro Giuria 1, I-10125 Torino, Italy}
\affiliation{Istituto Nazionale di Fisica Nucleare, Sezione di Torino, I-10125 Torino, Italy}
\affiliation{INAF-Osservatorio Astrofisico di Torino, via Osservatorio 20, 10025 Pino Torinese, Italy}

\author[0000-0003-3684-4275]{A. Capetti}
\affiliation{INAF-Osservatorio Astrofisico di Torino, via Osservatorio 20, 10025 Pino Torinese, Italy}

\author[0000-0002-9478-1682]{W. R. Forman}
\affiliation{Center for Astrophysics $|$ Harvard \& Smithsonian, 60 Garden Street, Cambridge, MA 02138, USA}

\author[0000-0002-0765-0511]{R. P. Kraft}
\affiliation{Center for Astrophysics $|$ Harvard \& Smithsonian, 60 Garden Street, Cambridge, MA 02138, USA}

\author[0000-0002-1824-0411]{R. D. Baldi}
\affiliation{Istituto di Radioastronomia, INAF, via Gobetti 101, 40129, Bologna, Italy}
\affiliation{Department of Physics \& Astronomy, University of Southampton, Hampshire SO17 1BJ, Southampton, UK}

\author[0000-0001-5221-2636]{V. H. Mahatma}
\affiliation{Th\"uringer Landessternwarte, Sternwarte 5, 07778 Tautenburg, Germany}

\author[0000-0002-5941-5214]{C. Mazzucchelli}
\affiliation{European Southern Observatory, Alonso de C\'ordova 3107, Vitacura, Regi\'on Metropolitana, Chile}

\author[0000-0001-8382-3229]{V. Missaglia}
\affiliation{Dipartimento di Fisica, Universit\`a degli Studi di Torino, via Pietro Giuria 1, I-10125 Torino, Italy}
\affiliation{Istituto Nazionale di Fisica Nucleare, Sezione di Torino, I-10125 Torino, Italy}
\affiliation{INAF-Osservatorio Astrofisico di Torino, via Osservatorio 20, 10025 Pino Torinese, Italy}

\author[0000-0001-5742-5980]{F. Ricci}
\affiliation{Dipartimento di Fisica e Astronomia dell’Universit\`a degli Studi di Bologna, via P. Gobetti 93/2, I-40129 Bologna, Italy}
\affiliation{INAF- Osservatorio di Astrofisica e Scienza dello Spazio di Bologna, via Gobetti 93/3, I-40129 Bologna, Italy}

\author[0000-0002-5941-5214]{G. Venturi}
\affiliation{Instituto de Astrof\'isica, Facultad de F\'isica, Pontificia Universidad Cat\'olica de Chile, Casilla 306, Santiago 22, Chile}
\affiliation{INAF - Osservatorio Astrofisico di Arcetri, Largo E. Fermi 5, I-50125 Firenze, Italy}

\author{S. A. Baum}
\affiliation{University of Manitoba,  Dept. of Physics and Astronomy, Winnipeg, MB R3T 2N2, Canada}

\author[0000-0003-0995-5201]{E. Liuzzo}
\affiliation{Istituto di Radioastronomia, INAF, via Gobetti 101, 40129, Bologna, Italy}

\author[0000-0001-6421-054X]{C. P. O'Dea}
\affiliation{University of Manitoba,  Dept. of Physics and Astronomy, Winnipeg, MB R3T 2N2, Canada}

\author[0000-0002-3585-2639]{M. A. Prieto}
\affiliation{Departamento de Astrof\'isica, Universidad de La Laguna, E-38206 La Laguna, Tenerife, Spain}
\affiliation{Instituto de Astrof\'isica de Canarias (IAC), E-38200 La Laguna, Tenerife, Spain}

\author[0000-0001-8887-2257]{H. J. A. R\"ottgering}
\affiliation{Leiden Observatory, Leiden University, PO Box 9513, 2300 RA Leiden, The Netherlands}

\author[0000-0002-3140-4070]{E. Sani}
\affiliation{European Southern Observatory, Alonso de C\'ordova 3107, Vitacura, Regi\'on Metropolitana, Chile}

\author[0000-0002-9011-6829]{W. B. Sparks}
\affiliation{SETI Institute, Mountain View, CA 94043}
\affiliation{Space Telescope Science Institute, 3700 San Martin Drive, Baltimore, MD 21218, USA}

\author[0000-0002-5445-5401]{G. R. Tremblay}
\affiliation{Center for Astrophysics $|$ Harvard \& Smithsonian, 60 Garden Street, Cambridge, MA 02138, USA}

\author[0000-0002-0587-1660]{R. J. van Weeren}
\affiliation{Leiden Observatory, Leiden University, PO Box 9513, 2300 RA Leiden, The Netherlands}

\author[0000-0003-1809-2364]{B. J. Wilkes}
\affiliation{Center for Astrophysics $|$ Harvard \& Smithsonian, 60 Garden Street, Cambridge, MA 02138, USA}

\author[0000-0003-0251-6126]{J. J. Harwood}
\affiliation{Centre for Astrophysics Research, School of Physics, Astronomy and Mathematics, University of Hertfordshire, College Lane,
Hatfield, Hertfordshire AL10 9AB, UK}

\author[0000-0002-5411-1748]{P. Mazzotta}
\affiliation{Dipartimento di Fisica, Universit\`a di Roma ``Tor Vergata” Via della Ricerca Scientifica 1, I-00133 Rome, Italy}

\author[0000-0001-5513-029X]{J. Kuraszkiewicz}
\affiliation{Center for Astrophysics $|$ Harvard \& Smithsonian, 60 Garden Street, Cambridge, MA 02138, USA}

\begin{abstract} 
We present the analysis of X-ray and optical observations of gas filaments observed in the radio source 3CR\,318.1, associated with NGC\,5920, the Brightest Cluster Galaxy (BCG) of MKW\,3s, a nearby cool core galaxy cluster. This work is one of the first X-ray and optical analyses of filaments in cool core clusters carried out using MUSE observations. We aim at identifying the main excitation processes responsible for the emission arising from these filaments. We complemented the optical VLT/MUSE observations, tracing the colder gas phase, with X-ray $\textit{Chandra}$ observations of the hotter highly ionized gas phase. Using the MUSE observations, we studied the emission line intensity ratios along the filaments to constrain the physical processes driving the excitation, and, using the $\textit{Chandra}$ observations, we carried out a spectral analysis of the gas along these filaments. We found a spatial association between the X-ray and optical morphology of these filaments, which are colder and have lower metal abundance than the surrounding intra-cluster medium (ICM), as already seen in other BCGs. Comparing with previous results from the literature for other BCGs, we propose that the excitation process that is most likely responsible for these filaments emission is a combination of star formation and shocks, with a likely contribution from self-ionizing, cooling ICM. Additionally, we conclude that the filaments most likely originated from AGN-driven outflows in the direction of the radio jet.

\end{abstract}


\section{Introduction}
\label{sec:intro}
The Third Cambridge Catalog of radio sources and its revised versions (3C, 3CR, 3CRR, \citealt{Edge1959,Bennett1962,Spinrad1985,Laing1983}) constitute one of the most valuable samples of radio-loud active galactic nuclei (AGN). In 2018, \citeauthor{Balmaverde2018} started the MUse RAdio Loud Emission line Snapshot (MURALES) survey to observe all southern 3CR sources with the Multi-Unit Spectroscopic Explorer (MUSE; \citealt{Bacon2010}) and enrich the multifrequency coverage of the 3CR catalog.

During the MURALES survey, an optical filamentary structure was detected in 3CR\,318.1 (\citealt{Balmaverde2019}), the brightest cluster galaxy (BCG) of the galaxy cluster MKW\,3s (\citealt{Morgan1975}), in agreement with observations by \citet{Edwards2009}. This structure consists of two ionized gas filaments extending $\sim$25 kpc in the south and the south-west directions, both with almost constant velocities along their projected lengths ($\sim$230 km s$^{-1}$ and $\sim$280 km s$^{-1}$, respectively). The filament that points to the south presents a bright knot at its southern end. In contrast with the typical narrow-line regions seen in other radio galaxies, these filaments extend beyond tens of kiloparsecs and present drastically different emission line ratios (e.g., [O III]/H$_\beta\sim$0.1 instead of the typical value of $\sim$10).

Similar filamentary structures were discovered in other BCGs, as NGC 1275 in the center of the Perseus galaxy cluster (\citealt{Lynds1970}, \citealt{Conselice2001} and \citealt{Fabian2008}), spatially associated with an X-ray excess \citep{Fabian2011}. The same situation occurs for other BCGs, where a tight connection between X-ray and optical filament emission was found \citep{Mcdonald2010}. 

However, the physical processes responsible for the ionization in these optical filaments is still debated. 
While \citet{Voit1994} proposed ionization due to hot cooling ICM as the excitation mechanism behind these filaments; \citet{Mcdonald2010}, who analyzed a sample of 23 filaments around BGCs, suggested that these types of filaments are due to ICM thermal conduction. In the case of the Perseus cluster, \citet{Fabian2011} rejected this scenario, due to the lack of a thick interface between the cold and hot gas, proposing that the origin of cold X-ray filaments is due to penetration of cold gas by the hot surrounding gas through reconnection diffusion. 

Here we present a comparison between optical and X-ray images of 3CR\,318.1 to investigate the main ionization processes underlying the origin of these optical filaments, thus focusing only on the first $\sim$30 kpc at the center of MKW\,3s (for works on the large scale structures see e.g., \citealt{Mazzotta2002,Mazzotta2004} and \citealt{Birzan2020}). 

This manuscript is organized as follows. A brief description of 3CR\,318.1 is given in \S~\ref{sec:target}. Optical and X-ray data analyses are reported in \S~\ref{sec:reduc}. Results are presented in \S~\ref{sec:results} while \S~\ref{sec:conclus} is devoted to our conclusions. 

Unless otherwise stated, we adopted cgs units for numerical results and assumed a flat cosmology with $H_0=69.6$ km s$^{-1}$ Mpc$^{-1}$, $\Omega_{M}=0.286$ and $\Omega_{\Lambda}=0.714$ \citep{bennett14}. Spectral indices, $\alpha$, are defined by flux density, S$_{\nu}\propto\nu^{-\alpha}$.

\section{3CR\,318.1}
\label{sec:target}
\begin{figure*}
   \centering
\includegraphics[width=18cm]{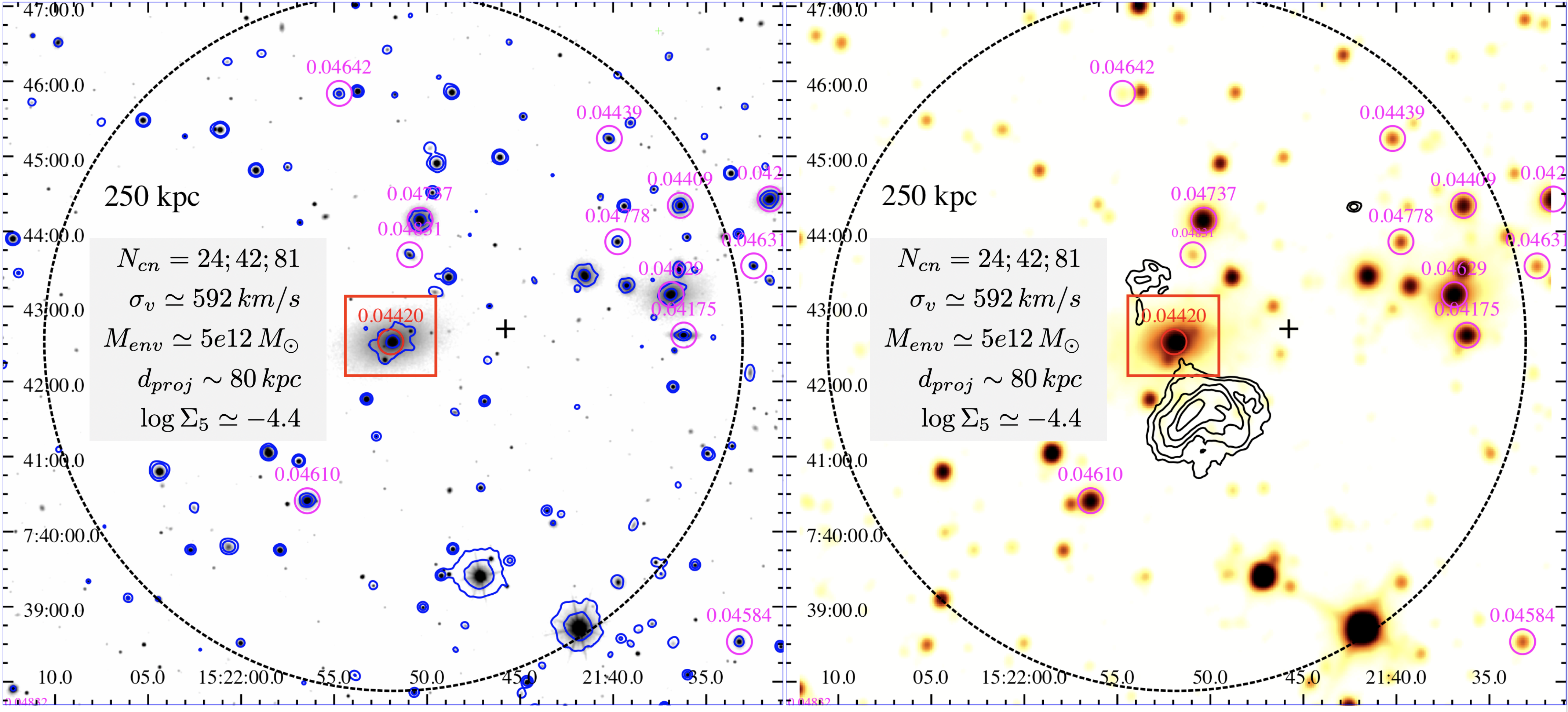}
\caption{r-band SDSS image with SWIFT UV contours at 2600 \AA\ overlaid in blue (left) and 3.4 $\mu$m WISE image with 328 MHz VLA radio contours overlaid in black (right), showing 3CR\,318.1 at the center and other sources in the field. The red squares show MUSE field of view. Radio contours were drawn at 5, 10, 20, 40, 80 times the root mean square (rms) level of the background in the radio map (beam size of 5.4\arcsec). Sources in the field with known spectroscopic redshifts are marked in magenta. Ambient parameters, shown in both panels, were derived according to \citet{Massaro2019,Massaro2020a,Massaro2020b}. $N_{cn}$ is the number of cosmological neighbors within a radius of 500\,kpc, 1\,Mpc and 2\,Mpc, respectively, $\sigma_v$ is the velocity dispersion of the cosmological neighbors, $M_{env}$ is the environmental mass (i.e., galaxies, IGM and dark matter) estimated from the velocity dispersion, $d_{proj}$ is the projected distance between the central radio galaxy and the centroid of the position of the cosmological neighbors within 2 Mpc (marked by the black cross) and $\Sigma_5$ is the cosmological parameter used to trace the dark matter halo density (see \citealt{Sabater2013} and \citealt{Worpel2013}). SWIFT-UVOT data were reduced according to the standard procedure (see \citealt{Massaro2008a,Massaro2008b}, for more details).}
\label{fig:overview}
\end{figure*}

3CR\,318.1 is the radio source associated with the nearby ($z=0.0453$, which corresponds to 0.896 kpc/arcsec) galaxy NGC\,5920, the BCG of the galaxy cluster MKW\,3s. Based on its optical line emission, it can be classified as an extremely low-excitation galaxy (see \citealt{Capetti2013}). The central source presents a steep spectrum at low radio frequencies (i.e., $\alpha^{150\text{ MHz}}_{1.4\text{ GHz}}=0.63$) with a luminosity of $\log L_{150\text{ MHz}}=23.16$ W Hz$^{-1}$ (\citealt{Capetti2020}). The surrounding diffuse radio emission also shows an extremely steep radio spectrum ($\alpha^{1.28\text{ GHz}}_{235\text{ MHz}}=2.42$), leading to a classification of relic radio galaxy by \citet{Giacintucci2007}. The host galaxy of 3CR\,318.1 presents an AB UV magnitude at 2600 \AA\ of 19.04 $\pm$ 0.04, corrected for Galactic reddening (\citealt{Cardelli1989}), with $E(B-V)=0.0356\pm0.0010$\footnote{https://irsa.ipac.caltech.edu/applications/DUST/}. This magnitude yields a UV luminosity at 2600 \AA\ of $L_{UV}=(4.96\pm0.17)\times10^{42}$ erg s$^{-1}$. An overview of 3CR\,318.1 and its environment is shown in Fig. \ref{fig:overview}.

\citet{Peres1998} found that MKW\,3s presents a moderate ``cooling flow" ($\dot{M} \sim$170 M$_\odot$ yr$^{-1}$). Using deep ($\sim$57 ks) $\textit{Chandra}$ X-ray observations of the central 200 kpc, \citet{Mazzotta2002} identified the presence of a cavity, located at $\sim$90 kpc south of the X-ray nucleus, hotter than the $\sim$3 keV average cluster temperature and filled by radio emission arising from the southern lobe (\citealt{Mazzotta2004}), as recently confirmed by \citet{Birzan2020} using Low Frequency Array observations. 

\section{Data reduction and analysis}
\label{sec:reduc}
X-ray data reduction was performed following standard procedures of the $Chandra$ Interactive Analysis of Observations v4.10 (CIAO; \citealt{Fruscione2006}) threads\footnote{http://cxc.harvard.edu/ciao/threads/}, adopting the \chn\ Calibration Database v4.8.4.1, and using the $\sim$57 ks $\textit{Chandra}$ observation (ObsID 900). The MUSE dataset was obtained from the MURALES survey (ID 099.B-0137(A); two exposures of 10 min each), with a mean seeing in the V band at zenith of 1.38\arcsec (see \citealt{Balmaverde2019} for details on the data reduction and analysis). Here we focus on the analysis of the [N II]$\lambda6584$ emission since it is more luminous and extended than the H$\alpha$ emission. The [N II]/H$\alpha$ ratio is constant along filaments with [N II]/H$\alpha\sim$1.9.

\begin{figure*}
    \centering
\includegraphics[width=16.cm]{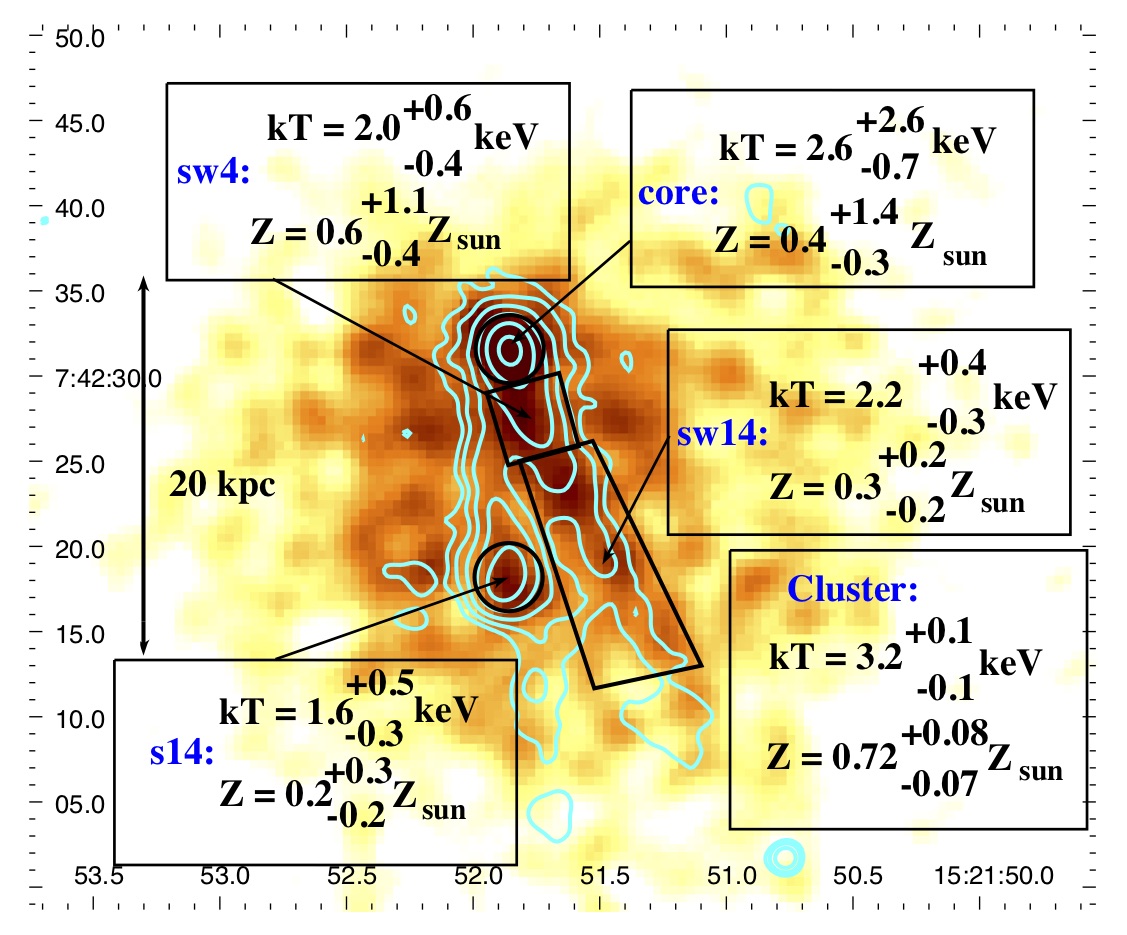}
\caption{0.5--3 keV $\textit{Chandra}$ image with [N II]$\lambda6584$ emission line contours overlaid in cyan. The $\textit{Chandra}$ image was smoothed with a 2\arcsec Gaussian kernel. [N II]$\lambda6584$ contours were drawn at 1, 2, 4, 8, 16 and 32 $\times10^{-20}$ erg s$^{-1}$ cm$^{-2}$ (starting at 3 times the rms). Regions chosen for the X-ray spectral analysis (described in \S~\ref{sec:reduc}) and the results for each region are shown in black. The fit for each region yielded the following reduced $\chi_\nu^2=1.304$ ($\nu=8$), with 237 net photons, for the core, $\chi_\nu^2=0.900$ ($\nu=5$), with 167 net photons, for region s14, $\chi_\nu^2=0.996$ ($\nu=10$), with 268 net photons, for region sw4, $\chi_\nu^2=0.588$ ($\nu=36$), with 845 net photons, for region sw14, and $\chi_\nu^2=0.906$ ($\nu=207$) for the cluster region.
}
\label{fig:spectral}
\end{figure*}

Astrometric registration to align radio, optical and X-ray images was performed by measuring the radio centroid at 1.4\,GHz (beam size of 4.5\arcsec), and the X-ray centroid in the 0.5--7 keV energy range and then aligning them following the procedure from \citet{Massaro2011}. We shifted the X-ray image 0.568\arcsec\ to the north-east, then, we aligned the position of the optical host galaxy in the MUSE continuum image with the radio centroid shifting it by 1.62\arcsec. These shifts are consistent with those previously reported for the 3CR \textit{Chandra} Snapshot Survey (\citealt{Massaro2011,Massaro2013,Massaro2015,Massaro2018,Jimenez2020}). Registration was then verified by checking the alignment of other point-like field sources in the Sloan Digital Sky Survey (SDSS) and in the MUSE continuum images. The comparison between registered X-ray and [NII]$\lambda6584$ optical images is shown in Fig. \ref{fig:spectral}.

In the same figure, we show the regions selected to perform X-ray spectral analysis, based on the morphology of the [N II]$\lambda6584$ emission. These, labeled according to the nomenclature of \citet{Massaro2011}, are:
\begin{enumerate}
    \item two circles of 2\arcsec\ radius, centered on (i) the radio core and (ii) the [N II]$\lambda6584$ emission line knot at the end of the south filament (core and s14, respectively);
    \item two polygons along the south-west filament, namely: sw4 (inner 4 kpc) and sw14 (remaining 13 kpc).
\end{enumerate}
We chose, as background for the spectral analysis of the filament regions, a larger circle located $\sim$45\arcsec\ to the north-east where no point-like and/or extended X-ray sources are detected, to obtain the spectrum of the excess emission only instead of the line-of-sight average. We also performed the spectral analysis of the galaxy cluster emission, considering two polygonal regions, one at each side of the filaments and a background region at $\sim$100\arcsec\ to compare the inner and outer galaxy cluster emission.
 
Background subtracted X-ray spectra in the 0.5--4 keV energy range were extracted using the CIAO routine \textsc{specextract} and the background region containing neighboring ICM emission to the filament and analyzed using Sherpa v4.12.1 (\citealt{Freeman2001}). To ensure the validity of Gaussian statistics, spectral data were binned to at least 25 photons per bin.

We tried fitting the final spectra using an absorbed bremsstrahlung model, however, it failed to reproduce the soft X-ray excess below 2 keV. Thus, we adopted a collisional ionisation gas model (\textsc{xsapec}\footnote{https://cxc.cfa.harvard.edu/sherpa/ahelp/xsapec.html}), absorbed by Galactic hydrogen column density (\textsc{xswabs}\footnote{https://cxc.cfa.harvard.edu/sherpa/ahelp/xswabs.html}; $N_{H,Gal}=3.8\times10^{20}$ cm$^{-2}$; \citealt{Kalberla2005}). The \textsc{xsapec} model had three free parameters: normalization, plasma temperature $kT$, and metal abundance $Z$.


We confirmed the presence of a soft X-ray (i.e., 0.5 - 3 keV) excess spatially coincident with the [N II]$\lambda6584$ filament (regions sw4 and sw14) at a level of confidence $>5\sigma$ with respect to the surrounding ICM emission. 
Detection significance, reported in Gaussian equivalent, was computed assuming a Poisson distribution of photons in the inner background region.

\begin{figure*}
    \centering
\includegraphics[width=5.9cm]{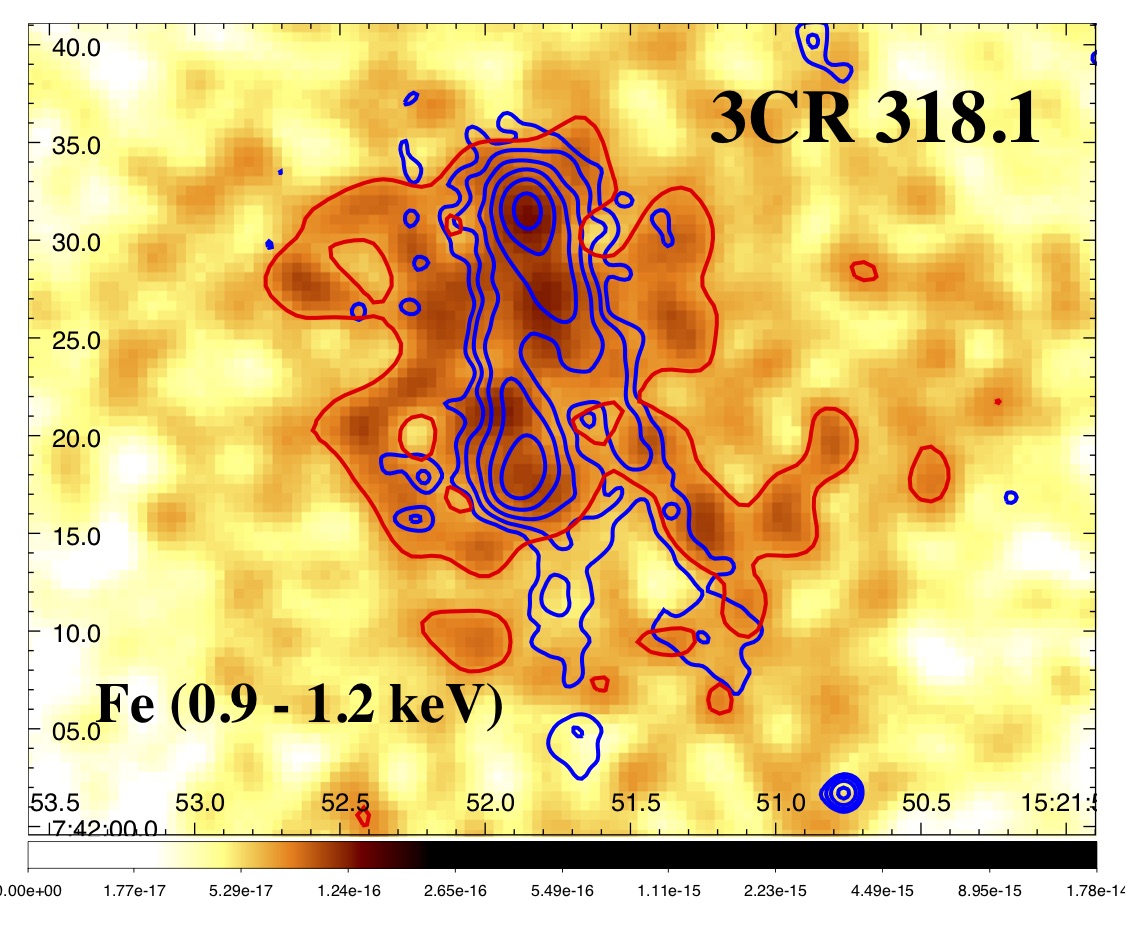}
\includegraphics[width=5.9cm]{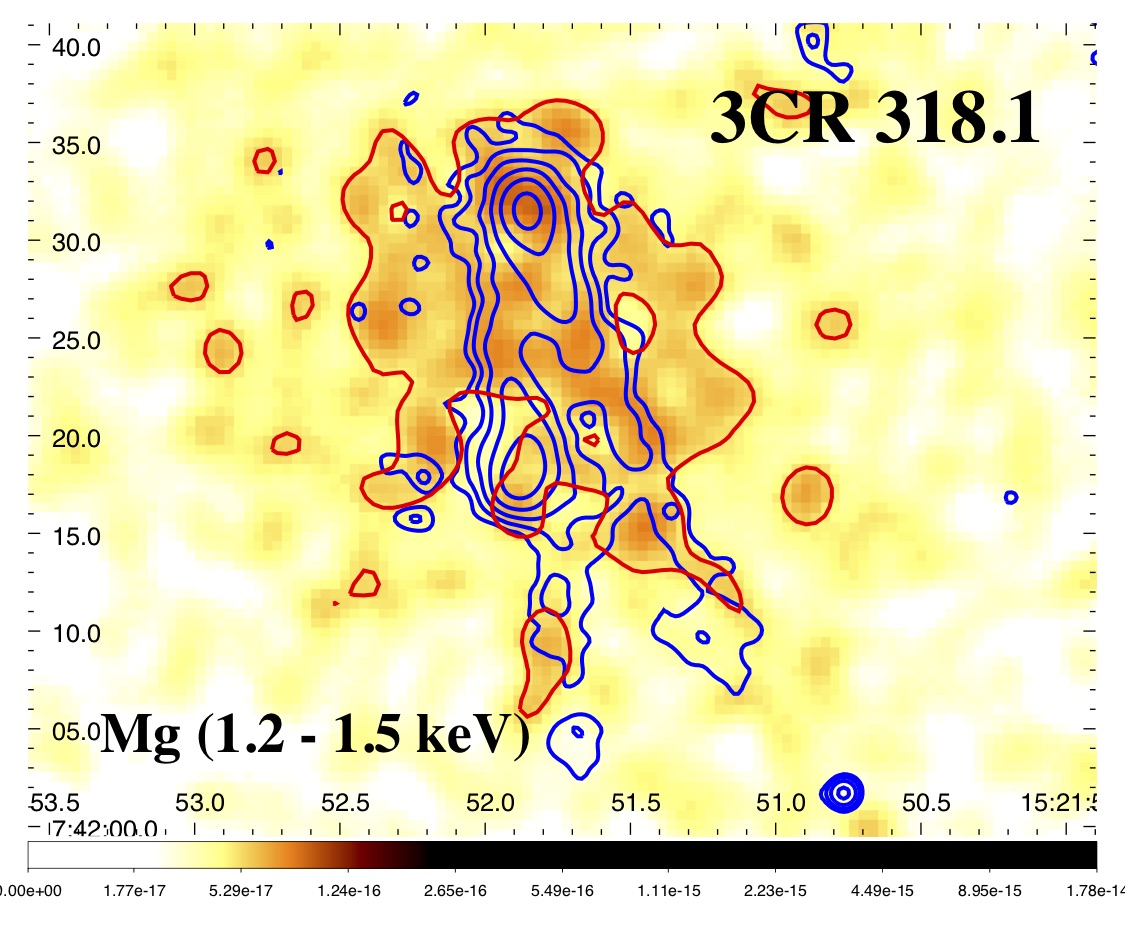}
\includegraphics[width=5.9cm]{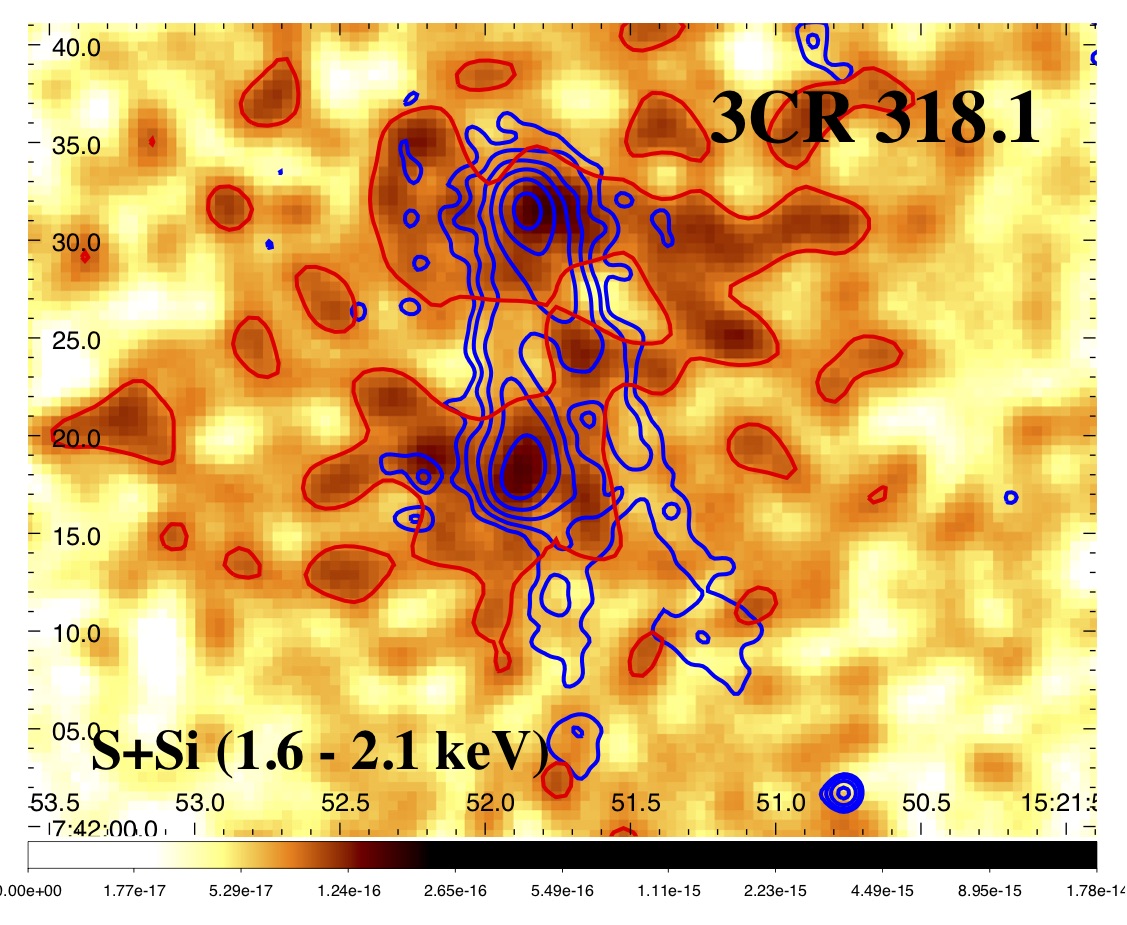}
\caption{Narrow-band $\textit{Chandra}$ flux images in the 0.9--1.2 keV (left), 1.2--1.5 keV (center) and 1.6--2.1 keV (right) energy bands to show Fe, Mg and S+Si X-ray emissions with [N II]$\lambda6584$ contours overlaid. \textit{Chandra} images were smoothed with a Gaussian kernel of 2\arcsec. [N II]$\lambda6584$ emission line contours are the same as in Fig. \ref{fig:spectral}. X-ray emission due to Mg seems to trace the [N II]$\lambda6584$ morphology the best. X-ray fluxes along the filament (regions sw4 and sw14) are $F_{\text{Fe}}=2.9^{+0.2}_{-0.3}\times10^{-14}$ erg s$^{-1}$ cm$^{-2}$ in the Fe band, $F_{\text{Mg}}=1.9^{+0.1}_{-0.1}\times10^{-14}$ erg s$^{-1}$ cm$^{-2}$ in the Mg band, and $F_{\text{S+Si}}=2.3^{+0.1}_{-0.2}\times10^{-14}$ erg s$^{-1}$ cm$^{-2}$ in the S+Si band.}
\label{fig:narrow}
\end{figure*}

Assuming the presence of emission lines below 2 keV (see \S~\ref{sec:results}), we created narrow band X-ray images, shown in Fig. \ref{fig:narrow} (see e.g., \citealt{Massaro2013,Massaro2015,Massaro2018}). We chose three energy ranges mainly attributable to (i) Fe (0.9--1.2 keV), (ii) Mg (1.2--1.5 keV) and (iii) S+Si (1.6--2.1 keV) ionized elements. X-ray flux maps were created by using monochromatic exposure maps set to the nominal energies of 1, 1.3 and 1.8 keV for the Fe, Mg, S+Si, respectively, to take into account the detector effective area at different energies.

From the optical perspective, we modeled the main emission lines present in the MUSE spectra, i.e., $H\beta$, $H\alpha$, [O III]$\lambda$5007, [O I]$\lambda$6300, [N II]$\lambda6584$ and the [SII] doublet at $\lambda\lambda$6716, 6731 using Gaussian functions at different locations across the filament and the knot. Top panel of Fig. \ref{fig:ratios} shows the MUSE spectrum of the optical knot sw14, as an example. The $H\beta$ line was modelled using an additional broad component only present in the core and in region s4.

\section{Results}
\label{sec:results}

\subsection{Optical}
\begin{figure*}
\centering
\includegraphics[width=8.9cm]{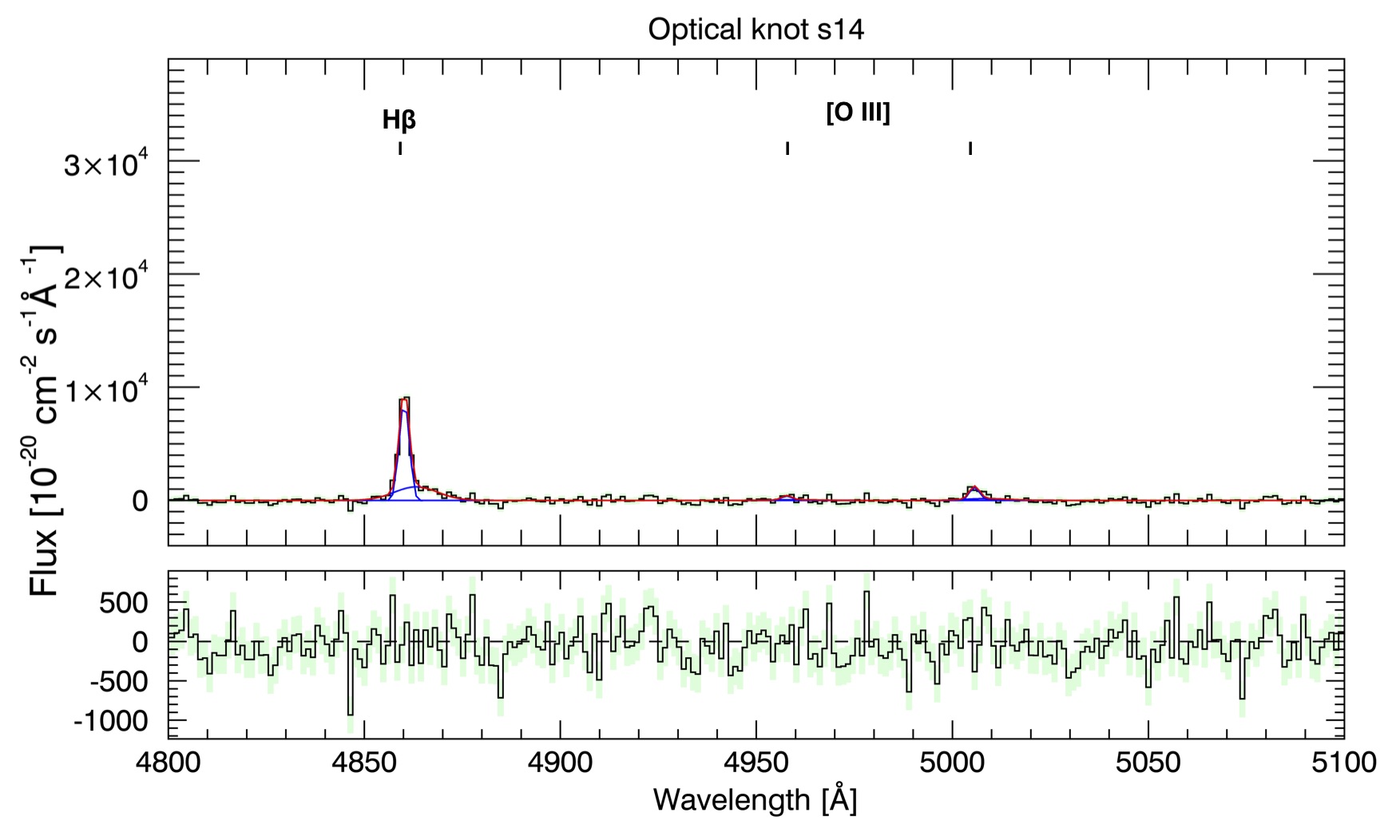}
\includegraphics[width=8.9cm]{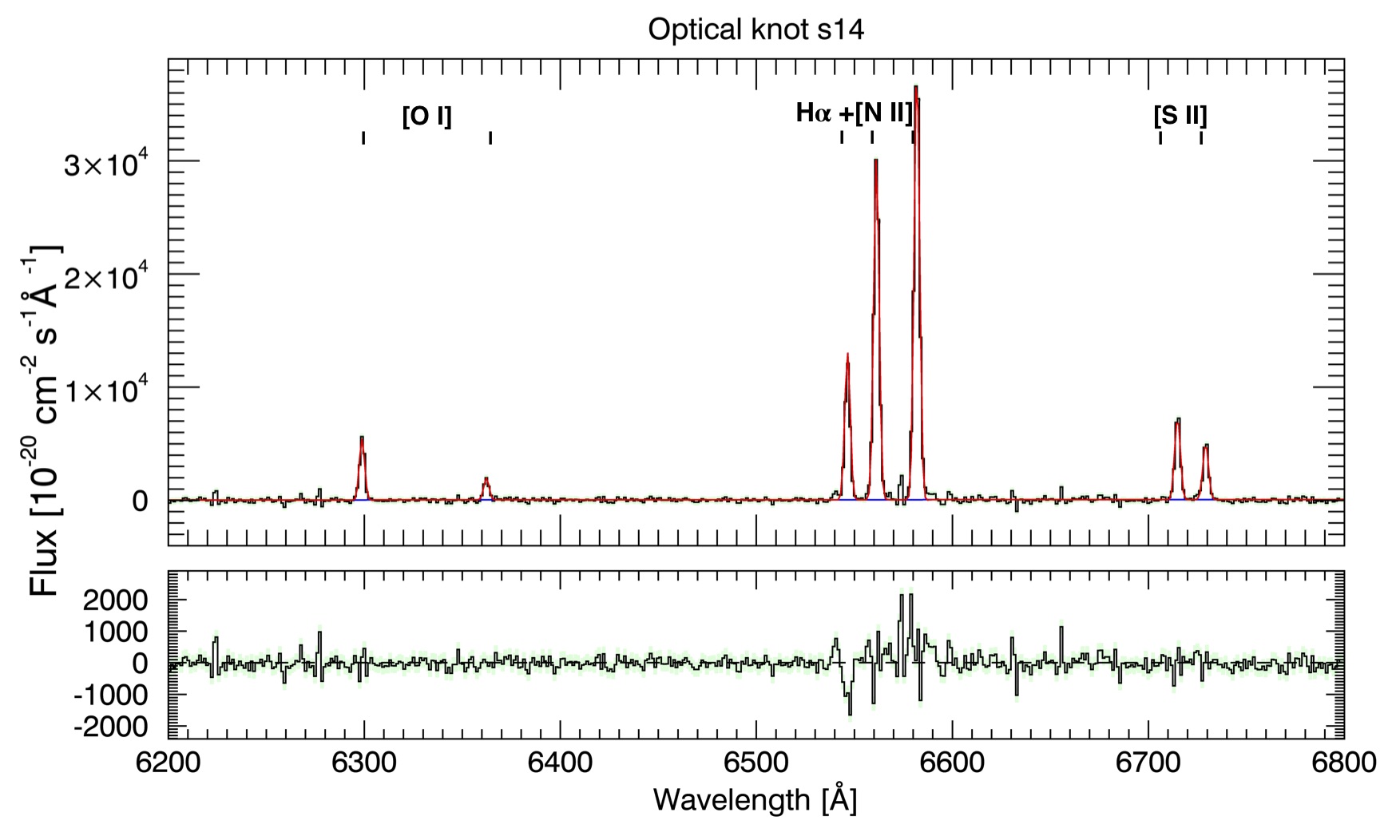}
\includegraphics[width=18.cm]{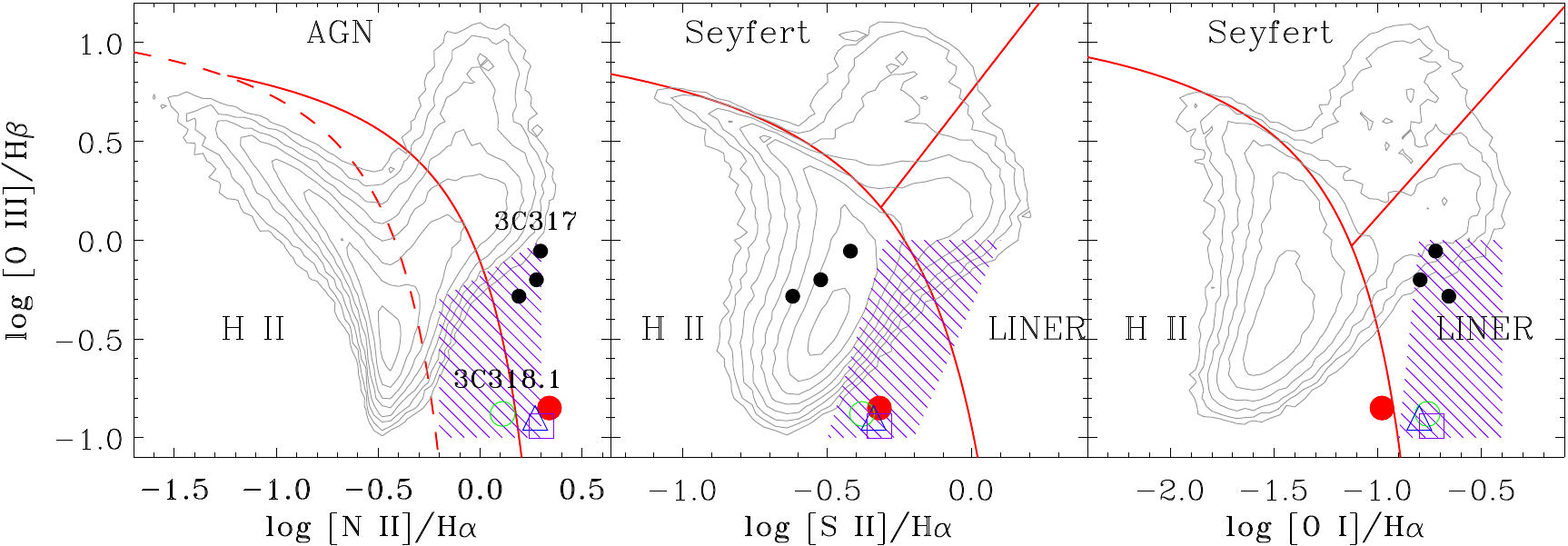}
\caption{
Top: MUSE spectrum (black histograms) of region s14, see Fig. \ref{fig:spectral}, in rest frame, shown as an example. We show in red the Gaussian fit of the emission lines (top panels) and the residuals (bottom panels).
Bottom: Location of 3CR\,318.1 regions ``core", s14, sw4 and sw14 (red filled circle, green circle, blue triangle and magenta square, respectively) in the BPT diagnostic diagrams. The red solid curves represent the \citet{Kewley2001} theoretical upper bound for pure star formation, the red straight line shows the \citet{Kewley2006} separation between AGN and LINERs, while the dashed red curve in the [N II]$\lambda6584$ BPT is the \citet{Kauffmann2003} empirical classification separating star-forming galaxies and AGN. Contours represent the iso-densities of all SDSS/DR7 emission line galaxies (\citealt{Capetti2011}). The three black dots mark the location of the filaments in 3CR\,317, the central galaxy of A2052, also observed as part of the MURALES survey (see \citealt{Balmaverde2018}). Purple dashed regions mark the location of filaments in other cool core galaxy clusters as measured by \citet{Mcdonald2012}.}
\label{fig:ratios}
\end{figure*}

We computed the emission line intensity ratios for all regions (see Fig. \ref{fig:spectral}) and plotted them in the diagnostic diagrams defined by the intensity ratios shown in Fig. \ref{fig:ratios}, used to distinguish between different ionization mechanisms occurring in H\,II regions, AGN, or LINER/shocks (\citealt{Baldwin1981} and \citealt{Dopita1995}). Measurements obtained for all four regions in 3CR\,318.1 are closely clustered and located where no emission line galaxies are found (see e.g., \citealt{Kewley2006} and \citealt{Capetti2011}). 

Similarly to what was found for filaments in different BCGs by \citet{Mcdonald2012}, line ratios provide a contradictory classification: at the boundary between LINER and H II from the [O III]/H$\beta$ vs [N II]/H$\alpha$ and [O I]/H$\alpha$ diagrams and in the region of ionization due to star-formation from the [S II]/H$\alpha$ ratio.


We also estimated the gas density along the filaments, using the ratio of the [S II]$\lambda$6716 and the [S II]$\lambda$6731 lines (see \citealt{Osterbrock1989}). Assuming a typical temperature of $10^4$ K, we obtained  $n_e\sim1380$ cm$^{-3}$ in the core and $n_e<100$ cm$^{-3}$ in s14 and along the filament (sw4 and sw14). We estimated the total mass of the ionized gas as $M=7.5\times10^{-3}\left(\frac{10^4}{n_e}\frac{L_{\text{H}_\beta}}{L_\odot}\right)$ M$_\odot$, where $L_{\text{H}_\beta}$ is the H$\beta$ luminosity (see \citealt{Osterbrock1989}). Thus, we estimated a total ionized gas mass of $M\sim2\times10^5$ M$_\odot$ in the core, $M>4\times10^5$ M$_\odot$ in s14, and $M>10^6$ M$_\odot$ along the filament.

Simulations carried out by \citet{Qiu2019,Qiu2020} show that cold filaments in cool core clusters can originate from warm AGN-driven outflows with shorter cooling than rising times. These simulations are consistent with the filaments in 3CR 318.1 extending in the direction of the southern radio lobe.

\subsection{X-rays}
We performed the X-ray spectral analysis in all regions described in \S~\ref{sec:reduc}, along optical filaments and in the galaxy cluster (excluding filaments, core and knots). Best fit results, obtained with the absorbed \textsc{xsapec} model, are reported in Fig. \ref{fig:spectral}. We also fitted regions sw4 and sw14 together and obtained a gas temperature of $2.4^{+0.3}_{-0.3}$ keV (with $Z=0.5^{+0.3}_{-0.2}$ $Z_\odot$ and reduced $\chi_\nu^2 =0.857$ with $\nu=59$ degrees of freedom) for the south-western filament. This is colder than the surrounding ICM, which has $kT=3.2^{+0.1}_{-0.1}$ keV (reduced $\chi_\nu^2 =0.906$; $\nu=207$). The ICM temperature is consistent with the average temperature found by \citet{Mazzotta2002} for the inner regions of MKW\,3s.

Metal abundance of the optical filament, $Z=0.3^{+0.2}_{-0.2}$ for sw14 (with reduced $\chi_\nu^2 =0.588$; $\nu=36$), appears to be marginally lower than that of the surrounding ICM ($Z=0.72^{+0.08}_{-0.07}$). This is in agreement with literature results on other filaments (see e.g., \citealt{Fabian2011} and \citealt{Mcdonald2010}).

Assuming that narrow band images trace the X-ray emission lines, the Mg emission has a better spatial association with the optical filaments, while Fe and S+Si show X-ray emission more extended than the filaments, as well as strong emission at the core and the  s14 (knot) regions (see Fig. \ref{fig:narrow}). We computed the X-ray flux along the filament (regions sw4 and sw14) in the Fe, Mg and S+Si band and compared them with optical and X-ray fluxes. The X-ray fluxes along the filament (regions sw4 and sw14) are $F_{\text{Fe}}=2.9^{+0.2}_{-0.3}\times10^{-14}$ erg s$^{-1}$ cm$^{-2}$ in the Fe band, $F_{\text{Mg}}=1.9^{+0.1}_{-0.1}\times10^{-14}$ erg s$^{-1}$ cm$^{-2}$ in the Mg band, $F_{\text{S+Si}}=2.3^{+0.1}_{-0.2}\times10^{-14}$ erg s$^{-1}$ cm$^{-2}$ in the S+Si band, and $F_{\text{0.5-3 keV}}=1.32^{+0.05}_{-0.08}\times10^{-13}$ erg s$^{-1}$ cm$^{-2}$ in the soft band (0.5 - 3 keV). Detection significance of the excess X-ray emission along optical filaments is above $5\sigma$ confidence level for both Mg and Fe, while it is not significant for the S+Si narrow band image, thus the S+Si emission along the filament is consistent with the ICM emission. 

We also obtained $F_{\text{H}\alpha}/F_{\text{Fe}}$ = 2.9$^{+0.4}_{-0.4}\times10^{-2}$ and $F_{\text{H}\alpha}/F_{\text{0.5-3 keV}}$ = 6.5$^{+0.7}_{-0.8}\times10^{-3}$. In contrast with the filaments of 3CR\,318.1, \citet{Sanders2007} and \citet{Fabian2011} found that those in Perseus have soft X-ray emission of the same order as the H$\alpha$ emission, while the H$\alpha$ emission is an order of magnitude above the Fe emission, which highlights the much lower ionization state here. Additionally, we obtained $F_{\text{[O III]}}/F_{\text{0.5-3 keV}}$ = 1.15$^{+0.12}_{-0.14}\times10^{-3}$, which is three orders of magnitude below the value obtained by \citet{Balmaverde2012} for the emission line regions in nine 3CR radio galaxies.

\begin{figure}
    \centering
\includegraphics[width=9cm]{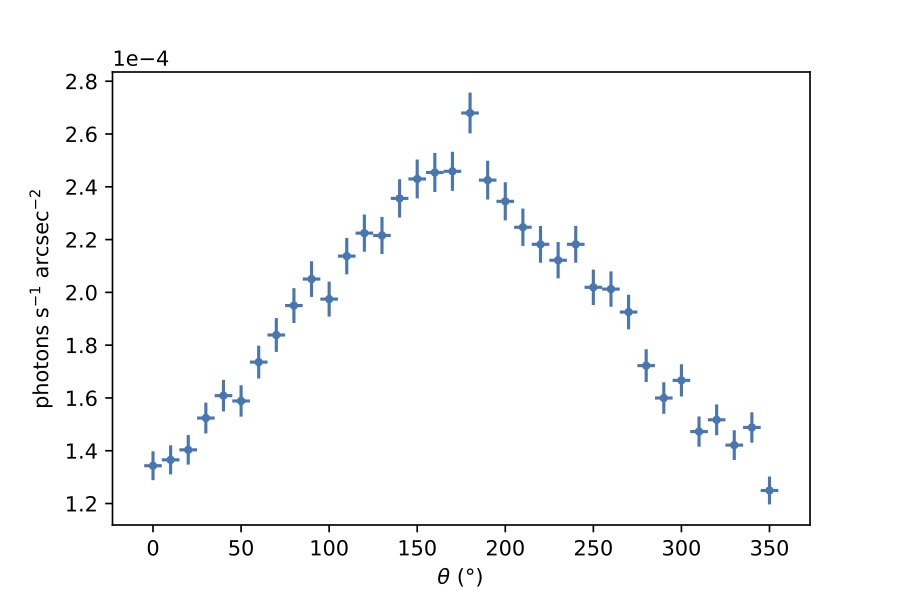}
\caption{0.5 - 3 keV, exposure corrected X-ray surface brightness profile with azimuthal bins of 10\degree\ centered on the core of 3CR\,318.1. The bins were chosen so that the bin at 180\degree\ is the one containing the [N II]$\lambda6584$ filament. Background level is two orders of magnitude below the filament emission. The increase of surface brightness corresponds to the [N II]$\lambda6584$.}
\label{fig:surfbrig}
\end{figure}

Lastly, we derived the 0.5--3 keV, exposure corrected, X-ray surface brightness profile with azimuthal bins of 10\degree\ centered on 3CR\,318.1 core, excluding the inner 2\arcsec\ and extending up to $\sim30$ kpc (see \citealt{Jimenez2021} for additional details). We chose to make the bin containing the [N II]$\lambda6584$ filament correspond to the 180\degree\ bin. The resulting X-ray surface brightness profile is shown in  Fig. \ref{fig:surfbrig}. The background level is two orders of magnitude below the filament emission. The smooth increase of the surface brightness towards the [N II]$\lambda6584$ filament supports the hypothesis that the filament could have originated due to outflows in the direction of the radio jets.

\section{Discussion and conclusions}
\label{sec:conclus}
The MURALES survey revealed the presence of  optical filaments in 3CR\,318.1, the BCG of MKW\,3s. Here we compared optical VLT/MUSE and X-ray \textit{Chandra} observations to shed light on their physical origin.

From an optical perspective, intensity ratios of various rest-frame emission lines were used to distinguish between different ionization mechanisms. Possible explanations of filamentary emission include: ionization due to hot cooling ICM (\citealt{Voit1994}), ICM thermal conduction (\citealt{Mcdonald2010}) and reconnection diffusion (\citealt{Fabian2011}). However, individually, none of the ionization mechanisms listed can account for all line ratios simultaneously. As shown in Fig. \ref{fig:ratios}, line ratios measured in 3CR\,318.1 show extremely low values of [O~III]/H$\beta$ in the diagnostic diagrams, similarly, \citet{Mcdonald2012} found systematically lower [O III]/H$\beta$ values in filaments of cool core galaxy clusters and higher values of [N II]/H$\alpha$, [O I]/H$\alpha$ and [S II]/H$\alpha$ than those found in galaxies in the SDSS. Although the diagnostic diagrams are used to discriminate between star-forming and AGN-dominated galaxies, they tend to fail when the ionization is due to more complex situations and/or different ionization mechanisms (see e.g., \citealt{Stasinska2008}, \citealt{Capetti2011} and \citealt{Balmaverde2018}) and therefore they may have a limited validity in assessing the ionisation conditions of BCGs. Thus, the line ratios in 3CR\,318.1 imply that the emission from the filaments is due to a combination of ionization mechanisms.  \citet{Mcdonald2012} suggested that line ratios are due to a combination of star formation and ionization from slow shocks ($\sim$100 - 400 km s$^{-1}$; see also \citealt{Allen2008}). Since our results are similar to those of \citet{Mcdonald2012}, we argue that their conclusion applies also to the case of 3CR\,318.1. An additional contribution from self-ionizing cooling ICM is suggested by the low velocity dispersion ($\sim 60$ km s$^{-1}$) of the optical filaments in 3CR\,318.1, which implies a small contribution of shocks to the total ionization. Therefore, we conclude that the underlying ionization mechanisms include a combination of photoionization due to star formation, self-ionizing cooling ICM, and a small contribution of ionization due to slow shocks. Additionally, although \citet{Mcdonald2012} already found a decrease of the emission line width with radius, thanks to the MUSE data, it was discovered that this decrease occurs sharply in the case of 3CR\,318.1 (from $\sim$200 km s$^{-1}$ in the core to $\sim$60 km s$^{-1}$ along the filaments; see \citealt{Balmaverde2019}).

We detected an excess of X-ray emission above the ICM along the [N II]$\lambda6584$ filaments. This X-ray filament is colder and despite the large uncertainties, appears to have a lower metal abundance than the surrounding ICM, in agreement with literature results (\citealt{Mcdonald2010}). The association between the X-ray and optical filamentary morphologies, together with the radio structure, suggests that the filaments could have originated from AGN-driven outflows in the direction of the radio jet. This scenario is in agreement with the smooth 0.5 - 3 keV X-ray surface brightness profile at the location of the [N II]$\lambda6584$ filament and with simulations carried out by \citet{Qiu2019,Qiu2020}, in which cold filaments originate from warm outflows (10$^4$ - 10$^7$ K). Additionally, works such as \citet{Gaspari2018} and \citet{Voit2021} predict that optical emission line nebulae would present velocity dispersions of $\sim$100 - 200 km s$^{-1}$ in the cases where the nebulae originated from compression and catastrophic cooling (see also \citealt{Gaspari2012, Gaspari2013, Gaspari2015, Gaspari2017} and \citealt{Voit2017} for previous works on multiphase condensation). However, the [N II]$\lambda6584$ filaments in 3CR\,318.1 present very low velocity dispersions ($\sim$60 km s$^{-1}$; see Fig. 8 in \citealt{Balmaverde2019}), so the AGN uplift scenario could be favoured.

\acknowledgments 
We thank the anonymous referee for their useful comments that led to the improvement of the paper. 
This work is supported by the ``Departments of Excellence 2018 - 2022’’ Grant awarded by the Italian Ministry of Education, University and Research (MIUR) (L. 232/2016). 
This research has made use of resources provided by the Ministry of Education, Universities and Research for the grant MASF\_FFABR\_17\_01. 
This investigation is supported by the National Aeronautics and Space Administration (NASA) grants GO9-20083X and GO0-21110X. 
F. M. is in debt with S. Bianchi for his valuable input on X-ray photoionizaton scenarios. 
A. J. acknowledges the financial support (MASF\_CONTR\_FIN\_18\_01) from the Italian National Institute of Astrophysics under the agreement with the Instituto de Astrofisica de Canarias for the ``Becas Internacionales para Licenciados y/o Graduados Convocatoria de 2017’’. 
A.P. acknowledges financial support from the Consorzio Interuniversitario per la fisica Spaziale (CIFS) under the agreement related to the grant MASF\_CONTR\_FIN\_18\_02.
W.F. and R.K. acknowledge support from the Smithsonian Institution and the Chandra High Resolution Camera Project through NASA contract NAS8-03060.
G.V. acknowledges support from ANID programs FONDECYT Postdoctorado 3200802 and Basal-CATA AFB-170002

\end{document}